\documentclass[runningheads]{svmult}
\usepackage{makeidx}   
\usepackage{graphicx}  
\usepackage{subeqnar}  
\usepackage{multicol}  
\usepackage{cropmark} 
\usepackage{physprbb}  
\begin{document}
\title*{The self-gravitating Fermi gas}
\toctitle{The self-gravitating Fermi gas}
%
%
\titlerunning{The self-gravitating Fermi gas}
%
\author{Pierre-Henri Chavanis}
\authorrunning{Pierre-Henri Chavanis}
%
%
\institute{Laboratoire de Physique Quantique, Universit\'{e} Paul Sabatier,
     118 route de Narbonne, 31062 Toulouse, France}

\maketitle              

\begin{abstract}

We discuss the nature of phase transitions in the self-gravitating
Fermi gas at non-zero temperature. This study can be relevant for
massive neutrinos in Dark Matter models and for collisionless
self-gravitating systems experiencing a ``violent relaxation''. Below
a critical energy (in the microcanonical ensemble) or below a critical
temperature (in the canonical ensemble), the system undergoes a
gravitational collapse leading to a compact object with a massive
degenerate nucleus (fermion ball).

\end{abstract}

\section{Introduction}

The self-gravitating Fermi gas model was introduced by Fowler in 1926
in order to explain the puzzling nature of white dwarf stars. When a
star has exhausted its nuclear fuel, it undergoes a gravitational
collapse until gravity is balanced by the pressure of a degenerate gas
of electrons (on account of Pauli's exclusion principle). This model
was later on improved by Chandrasekhar (1930) who took into account
special relativity effects in the degenerate equation of state of
electrons and discovered that white dwarf stars have a limiting mass
of $\sim 1.4\ M_{\odot}$. Then, in 1939, Oppenheimer \& Volkoff
considered the extension of this model to the framework of general
relativity in connexion with the structure of neutron stars. They
found again a limiting mass of $\sim 0.7\ M_{\odot}$ above which the
pressure of the neutrons cannot sustain gravity anymore. In that case,
the star is expected to collapse into a black hole. In these studies,
the Fermi gas is completely degenerate since the thermal energy $kT$
is much smaller than the Fermi energy. The self-gravitating Fermi gas
at finite temperature was investigated by Hertel \& Thirring
(1971). They showed that below a critical temperature a homogeneous
gas of fermions undergoes a first order phase transition leading to a
compact object with a degenerate core. Application of these results to
Dark Matter was considered by Bilic \& Viollier (1997). They proposed
that Dark Matter could be made of massive neutrinos (with $m\sim 15
{\rm keV}$) in equilibrium with a radiation background imposing its
temperature. In the langage of statistical mechanics, this corresponds
to the {\it canonical ensemble}. By cooling below a critical
temperature, a condensed phase emerges consisting of a quasidegenerate
``fermion ball''. These fermion balls may provide an alternative to
black holes that are reported to exist at the center of galaxies. At
large distances, the $r^{-2}$ law of density decrease of an isothermal
gas is consistent with the flat rotation curves of spiral galaxies.
This could provide an attractive model of Dark Matter but the
dynamical mechanism leading to a ``fermion ball'' remains to be
clearly identified.

The self-gravitating Fermi gas model also appeared in a completely
different context, independant of quantum mechanics. In 1967,
Lynden-Bell argued that a self-gravitating system far from mechanical
equilibrium would rapidly relax towards a virialized state, due to the
strong fluctuations of the gravitational potential. Since this process
of {\it violent relaxation} is essentially collisionless, the
coarse-grained distribution function $\overline{f}$ cannot increase
and it must satisfy an effective exclusion principle $\overline{f}\le
\eta_{0}$, where $\eta_{0}$ is the maximum value of the initial
distribution function. This upper bound is a consequence of the
Liouville theorem. Assuming ergodicity, Lynden-Bell predicted that
$\overline{f}$ should converge towards a Fermi-Dirac distribution, or
a superposition of Fermi-Dirac distributions. Relaxation equations
towards these maximum entropy states (on a coarse-grained scale) were
proposed by Chavanis, Sommeria and Robert (1996). In that context, the
proper thermodynamical ensemble is the {\it microcanonical ensemble}
since energy is conserved during the course of the evolution.
Fermi-Dirac spheres were computed by Chavanis \& Sommeria (1997). They
found a wide variety of nuclear concentration depending on the degree
of degeneracy and on the value of energy. For high energies, the
system is almost uniform. For intermediate energies, the core is
partially degenerate. For low energies, the maximum entropy (i.e. most
probable) state consists of a massive degenerate nucleus surrounded by
an isothermal halo. They argued that degeneracy (in Lynden-Bell's
sense) could stabilize the system and play a role in galactic nuclei
and Dark Matter. It is however unclear whether violent relaxation can
lead to {\it massive} ``fermion balls''. Indeed, it is in general advocated
that the fluctuations of the gravitational potential fade before the
system has developed high density contrasts. To our point of view,
this remains a matter for further investigation.

Recently, we have carried on a detailed study of phase transitions in
the self-gravitating Fermi gas (Chavanis 2002). We showed that this
system exhibits a rich structure with the occurence of three types of
phase transitions of zeroth, first and second order. We worked at a
general level without specifying the source of degeneracy (quantum
mechanics or violent relaxation) and we described the complete
structure of the equilibrium phase diagram, for arbitrary values of control
parameters and arbitrary degree of degeneracy. We worked both in the
microcanonical and canonical ensembles, emphasizing the inequivalence
of these ensembles for long-range systems. The description of the
microcanonical ensemble and the relation between the structure of
Fermi-Dirac spheres and classical isothermal spheres investigated by
Antonov (1962) and Lynden-Bell \& Wood (1968) is a specificity of our
approach. In the following, we give a short summary of this study.

\section{Phase transitions in self-gravitating systems}

We consider a system of $N$ fermions of mass $m$ interacting via Newtonian gravity. At statistical equilibrium, the system is described by the Fermi-Dirac distribution
\begin{equation}
{f}={\eta_{0}\over 1+\lambda e^{\beta ({v^{2}\over 2}+\Phi)}}, \qquad (\beta={m/ kT})
\label{fFD}
\end{equation}
coupled to the Poisson equation
\begin{equation}
\Delta \Phi =4\pi G \int f d^{3}{\bf v}.
\label{Mf}
\end{equation}
This determines a self-consistent meanfield equation for the gravitational potential $\Phi$. For spherically symmetrical systems, this is just an ordinary differential equation, which can be solved numerically by usual means. Then, the fugacity $\lambda^{-1}>0$ and the inverse temperature  $\beta$ can be related to the total mass $M$ and total energy $E$ of the system. In Eq. (\ref{fFD}), $\eta_{0}$ represents the maximum allowable value of the distribution function, i.e. $f\le \eta_{0}$. For a quantum gas, $\eta_{0}=(2s+1)m^{4}/h^{3}$ where $s$ is the spin of the particles. In the fully degenerate limit $f\simeq \eta_{0}$, the system is equivalent to a polytrope of index $n=3/2$.  In the non-degenerate limit $f\ll\eta_{0}$, Eqs. (\ref{fFD})-(\ref{Mf}) describe a classical isothermal gas  
\begin{equation}
{f}={\eta_{0}\over\lambda} e^{-\beta ({v^{2}\over 2}+\Phi)}\qquad {\rm and}\qquad \Delta\Phi=4\pi GA e^{-\beta\Phi}.
\label{PV6}
\end{equation} 
The Boltzmann-Poisson system (\ref{PV6}) has been studied in relation
with the structure of isothermal stellar cores and globular clusters.  It is
well-known that the density of an isothermal gas decreases at large
distances like $r^{-2}$. Therefore, the total mass of the
configuration is infinite and we need to introduce truncated
models (Chavanis 1998) or confine the system within a box of radius $R$.

\begin{figure}
\begin{center}
\includegraphics[width=0.7\textwidth]{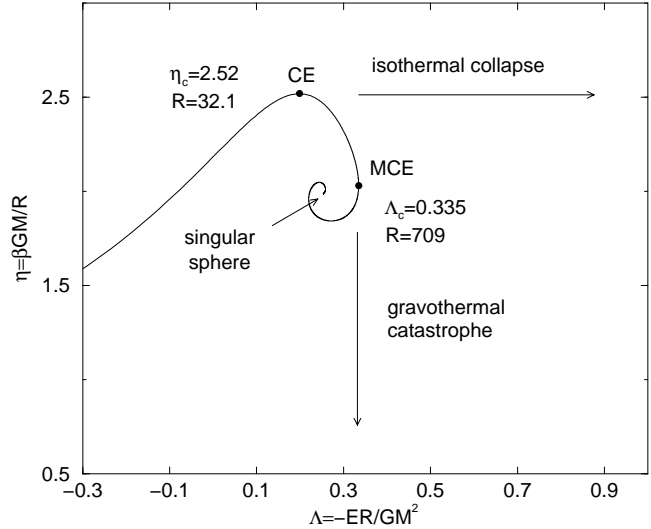}
\end{center}
\caption[]{Equilibrium phase diagram for classical isothermal spheres. For $\Lambda>\Lambda_{c}$ or $\eta>\eta_{c}$, there is no hydrostatic equilibrium and the system undergoes a gravitational collapse.}
\label{etalambda}
\end{figure}

\begin{figure}
\begin{center}
\includegraphics[width=0.7\textwidth]{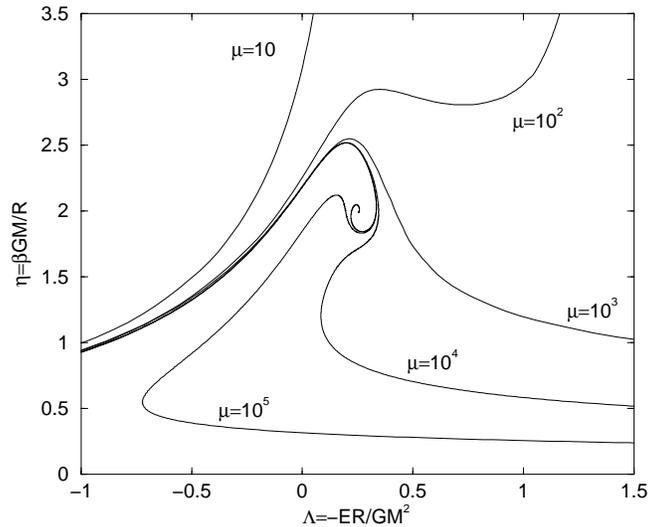}
\end{center}
\caption[]{Equilibrium phase diagram for self-gravitating fermions with different values of degeneracy parameter $\mu$. For $\mu\rightarrow +\infty$, the spiral makes several rotations before unwinding \cite{chav1,ispolatov}. }
\label{multimu}
\end{figure}

In the non degenerate limit, the equilibrium phase diagram $(E,T)$ is
represented in Fig. \ref{etalambda}. The curve has a striking spiral
behaviour parametrized by the density contrast ${\cal
R}=\rho(0)/\rho(R)$ going from $1$ (homogeneous system) to $+\infty$
(singular sphere) as we proceed along the spiral. There is no
equilibrium state below $E_{c}=-0.335GM^{2}/R$ or $T_{c}={GMm\over 2.52
kR}$. In that case, the system is expected to collapse
indefinitely. This is called {\it gravothermal catastrophe} in the
microcanonical ensemble (fixed $E$) and {\it isothermal collapse} in
the canonical ensemble (fixed $T$). Dynamical models show that the
collapse is self-similar and develops a finite time singularity (see
Chavanis {\it et al.} 2002 and references therein). However,
although the central density goes to $+\infty$, the shrinking of the
core is so rapid that the core mass goes to zero. Therefore, the
singularity contains no mass and this process alone cannot lead to a
black hole.

It is also important to stress that the statistical ensembles are not
interchangeable for systems with long-range interaction, like
gravity. In the microcanonical ensemble, the series of equilibria
becomes unstable after the first turning point of energy $(MCE)$
corresponding to a density contrast of $709$. At that point, the
solutions pass from local entropy maxima to saddle points.
In the canonical ensemble, the series of equilibria becomes unstable
after the first turning point of temperature $(CE)$ corresponding to a
density contrast of $32.1$. At that point, the solutions
pass from minima of free energy ($F=E-TS$) to saddle points.
It can be noted that the region of negative specific heats between
$(CE)$ and $(MCE)$ is stable in the microcanonical ensemble but
unstable in the canonical ensemble as expected on general physical grounds.

\begin{figure}
\begin{center}
\includegraphics[width=0.7\textwidth]{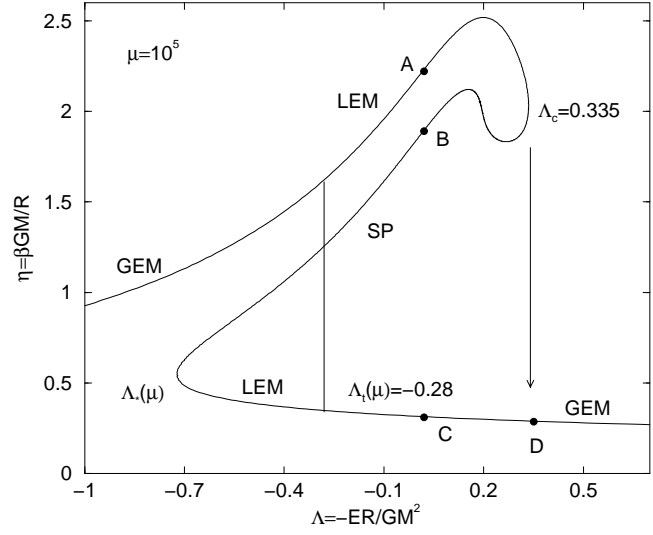}
\end{center}
\caption[]{Equilibrium phase diagram for Fermi-Dirac spheres with a degeneracy parameter $\mu=10^{5}$. Points A  form the ``gaseous''  phase. They are global entropy maxima (GEM) for $\Lambda<\Lambda_{t}(\mu)$ and local entropy maxima (LEM), i.e. {\it metastable states}, for $\Lambda>\Lambda_{t}(\mu)$. Points C form the ``condensed'' phase. They are LEM for $\Lambda<\Lambda_{t}(\mu)$ and  GEM for $\Lambda>\Lambda_{t}(\mu)$. Points B are unstable saddle points (SP) and contain a ``germ''.}
\label{le5}
\end{figure}
\begin{figure}
\begin{center}
\includegraphics[width=0.7\textwidth]{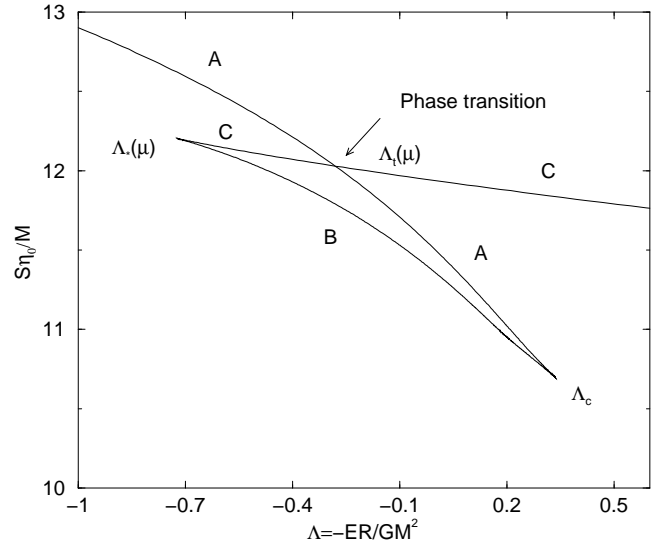}
\end{center}
\caption[]{Entropy of each phase versus energy for $\mu=10^{5}$. A first order phase transition is expected at $\Lambda_{t}(\mu)$ at which the two stable branches (solutions A and C) intersect. However, the entropic barrier $B$ probably prevents this phase transition \cite{chav1,ispolatov}. }
\label{SL5}
\end{figure}

When degeneracy is taken into account, the structure of the
equilibrium phase diagram depends on the value of the degeneracy
parameter $\mu=\eta_{0}\sqrt{512\pi^{4}G^{3}MR^{3}}$. The classical
limit is recovered for $\mu\rightarrow +\infty$. We see that the
inclusion of degeneracy has the effect of unwinding the spiral
(Fig. \ref{multimu}). For large values of the degeneracy parameter,
the equilibrium phase diagram is depicted in Fig. \ref{le5}. The solutions
on the upper branch (points A) are non degenerate and have a smooth
density profile; they form the ``gaseous phase''. The solutions on the
lower branch (points $C$) have a ``core-halo'' structure with a
massive degenerate nucleus and a dilute atmosphere; they form the
``condensed phase''. The density profiles of these solutions are given
in Ref. \cite{chav1}. By comparing their entropy (Fig. \ref{SL5}), we
would expect that a first order phase transition from the gaseous
phase to the condensed phase occurs at the transition energy
$E_{t}(\mu)$. This is, however, unlikely because the probability of a
fluctuation able to induce this phase transition is extremely weak
\cite{ispolatov}. Therefore, the {\it metastable} gaseous states with
$E<E_{t}$ are probably physical for the time scales involved in astrophysical
situations. In any case, a phase transition {\it must} occur at the
critical energy $E_{c}$ at which the gaseous branch disappears. Below
that energy, the system undergoes a gravothermal catastrophe but, for
self-gravitating fermions, the core ceases to shrink when it becomes
degenerate. Since this collapse is accompanied by a discontinuous jump
of entropy, this is sometimes called a zeroth order phase
transition. The resulting equilibrium state (point $D$) possesses a
small degenerate nucleus which contains a moderate
fraction of the total mass ($\simeq 20\%$ for $\mu=10^{5}$). The rest of the
mass is diluted in a hot envelope held by the box. In an open system,
it would be dispersed at infinity.

\begin{figure}
\begin{center}
\includegraphics[width=0.7\textwidth]{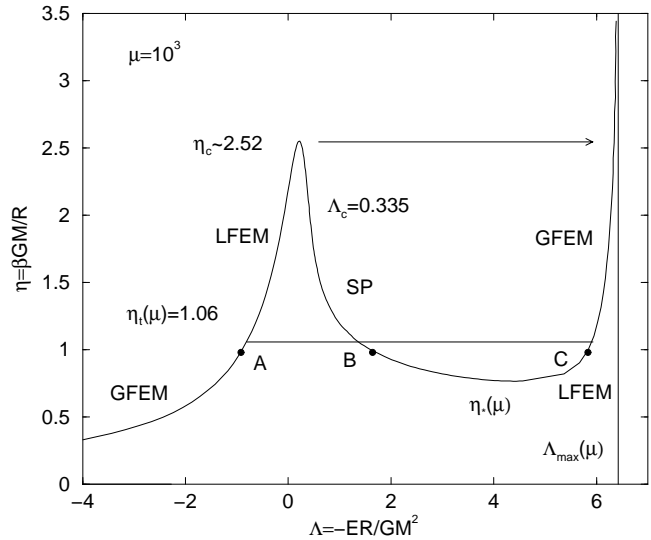}
\end{center}
\caption[]{Equilibrium phase diagram for Fermi-Dirac spheres with $\mu=10^{3}$.}
\label{fel}
\end{figure}

\begin{figure}
\begin{center}
\includegraphics[width=0.7\textwidth]{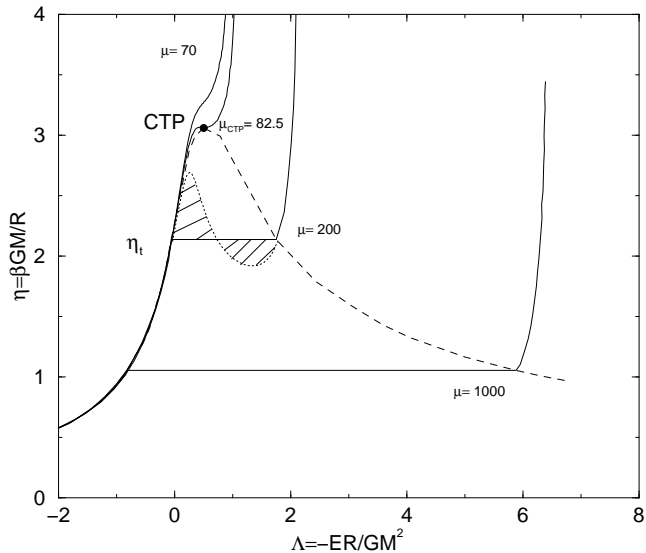}
\end{center}
\caption[]{Enlargement of the phase diagram near the
tricritical point in the canonical ensemble. The Maxwell construction
determining the transition temperature $\eta_{t}(\mu)$ is made
explicitly (dashed areas). For $\mu_{CTP}=82.5$ the Maxwell plateau
disappears and the $\Lambda-\eta$ curve presents an inflexion point at
$\Lambda_{CTP}\simeq 0.5$, $\eta_{CTP}\simeq 3.06$. At that point, the
specific heat becomes infinite and the transition is second
order. This diagram is remarkably similar to the liquid/gas transition
for an ordinary fluid. }
\label{tricritique}
\end{figure}
\begin{figure}
\begin{center}
\includegraphics[width=0.7\textwidth]{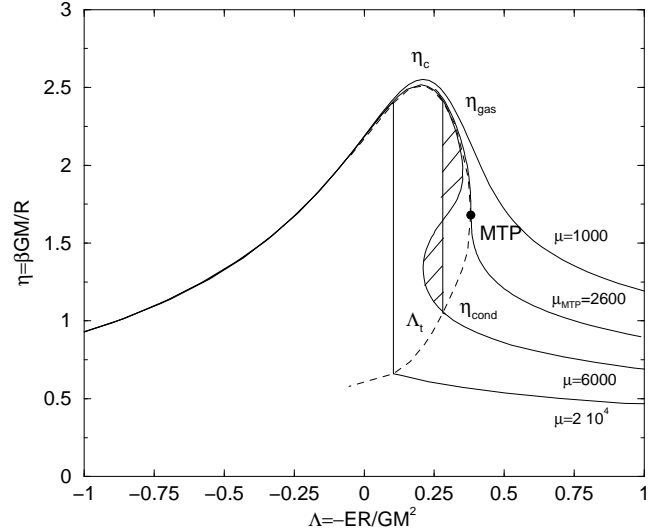}
\end{center}
\caption[]{Same as Fig. \ref{tricritique} near the tricritical point in the microcanonical ensemble ($\mu_{MTP}\simeq 2600$, $\Lambda_{MTP}\simeq 0.38$, $\eta_{MTP}\simeq 1.68$).}
\label{tricritiquemicro}
\end{figure}

For smaller values of the degeneracy parameter, the equilibrium phase
diagram is represented in Fig. \ref{fel}. The curve $\eta(\Lambda)$ is
now univalued and the first order phase transition in the
microcanonical ensemble is suppressed: all the equilibrium states are
global entropy maxima. For large energies, they are almost homogeneous
and for smaller energies they have a ``core-halo'' structure with a
partially degenerate nucleus. As energy is progressively decreased,
the nucleus becomes more and more degenerate and contains more and
more mass until a minimum energy $E_{min}=-2.36\
G^{2}M^{7/3}\eta_{0}^{2/3}$ at which all the mass is in the completely
degenerate nucleus of radius
$R_{*}=0.181\ \eta_{0}^{-2/3}G^{-1}M^{-1/3}$. In that case, the system
has the same structure as a cold white dwarf star. Therefore,
depending on the degree of degeneracy and on the value of energy, a
wide variety of nuclear concentrations can be obtained in the Fermi gas.

We can also consider the situation in which the system is in contact
with a heat bath which imposes its temperature (canonical
description). Considering again the case $\mu=10^{3}$ in
Fig. \ref{fel}, we note that the curve $\Lambda(\eta)$ is
multivalued. We expect therefore a first order phase transition to
occur at a transition temperature $T_{t}$ determined by a Maxwell
construction. This phase transition should be accompanied by a huge
release of latent heat. This may not be physically realizable and the
true collapse will rather occur at $T_{c}$. The outcome of this
collapse is the formation of a fermion ball containing almost all the
mass. Considering Fig. \ref{multimu} again, we see that the phase
transitions are suppressed in the microcanonical ensemble for
$\mu<\mu_{MTP}\simeq 2600$ and in the canonical ensemble for
$\mu<\mu_{CTP}\simeq 82.5$. At these {\it tricritical points}, the two
phases (gaseous and condensed) merge. This characterizes a second
order phase transition (see
Figs. \ref{tricritique}-\ref{tricritiquemicro}). The analogies and the
differences with the liquid/gas transition are described in Ref. \cite{chav2}.

%

\end{document}